\documentclass{article}
\usepackage{moreverb}
\usepackage{amsfonts}
\usepackage{graphicx}
\usepackage{algorithm}
\usepackage[end]{algpseudocode}
\usepackage[dvips,colorlinks,bookmarksopen,bookmarksnumbered,citecolor=red,urlcolor=red]{hyperref}

\newcommand\naive{na{\"\i}ve}

\algnewcommand{\algorithmicgoto}{\textbf{go to}}
\algnewcommand{\Goto}[1]{\algorithmicgoto~\ref{#1}}

\begin{document}

\title{Technology Beats Algorithms (in Exact String Matching)\footnote{Draft of the paper submitted to \textit{Software: Practice \&{} Experience}}}

\author{Jorma Tarhio\footnote{Department of Computer Science, Aalto University, Finland}, Jan Holub\footnote{Department of Theoretical Computer Science,
Faculty of Information Technology,
Czech Technical University in Prague, Czech Republic,
\texttt{Jan.Holub@fit.cvut.cz}}, and Emanuele Giaquinta\footnote{F-Secure Corporation, Finland}
}

\maketitle

\begin{abstract}
  More than 120 algorithms have been developed for exact string matching within the last 40 years. We show by experiments that the \naive{} algorithm exploiting SIMD instructions of modern CPUs (with symbols compared in a special order) is the fastest one for patterns of length up to about 50 symbols and extremely good for longer patterns and small alphabets. The algorithm compares 16 or 32 characters in parallel by applying SSE2 or AVX2 instructions, respectively. Moreover, it uses loop peeling to further speed up the searching phase. We tried several orders for comparisons of pattern symbols and the increasing order of their probabilities in the text was the best. 
\end{abstract}


\section{Introduction}

The exact string matching is one of the oldest tasks in computer science. The need for it started when computers began processing text. At that time the documents were short and there were not so many of them. Now, we are overwhelmed by amount of data of various kind. The string matching is a crucial task in finding information and its speed is extremely important.

The exact string matching task is defined as counting or reporting all the locations of given pattern $p$ of length $m=|p|$ in given text $t$ of length $n=|t|$ assuming $m \ll n$, where $p$ and $t$ are strings over a finite alphabet $\Sigma$. The first solutions designed were to build and run deterministic finite automaton \cite{AHU74} (running in space ${\cal O}(m|\Sigma|)$ and time ${\cal O}(n)$), the Knuth--Morris--Pratt automaton \cite{KMP77} (running in space ${\cal O}(m)$ and time ${\cal O}(n)$), and the Boyer--Moore algorithm \cite{BM77} (running in best case time $\Omega(n/m)$ and worst case time ${\cal O}(mn)$). There are numerous variations of the Boyer--Moore algorithm like \cite{Hor80,ZT87,Sun90,HS91}. In total more than 120 exact string matching algorithms \cite{Far2016} have been developed since 1970.

  Modern processors allow computation on vectors of length 16 bytes in case of SSE2 and 32 bytes in case of AVX2. The instructions operate on such vectors stored in special registers XMM0--XMM15 (SSE2) and YMM0--YMM15 (AVX2). As one instruction is performed on all data in these long vectors, it is considered as SIMD (Single Instruction, Multiple Data) computation.

\section{Algorithms}

\subsection{Na{\"\i}ve Approach}

In the \naive{} approach (shown as Algorithm~\ref{NaiveSearch2}) the pattern $p$ is checked against each position in the text $t$ which leads to running time ${\cal O}(mn)$ and space ${\cal O}(1)$. However, it is not bad in practice for large alphabets as it performs only 1.08 comparisons \cite{HS91} on average on each character of $t$ for English text. The variable \textit{found} in Algorithm~\ref{NaiveSearch2} is not quite necessary. It is presented in order to have a connection to the SIMD version to be introduced.

Like in the testing evironment of Hume \&{} Sunday \cite{HS91} and the SMART library \cite{FLBDM2016}, we consider the counting version of exact string matching. It can be is easily transformed into the reporting version by printing position $i$ in line~\ref{NaiveSearch2-printi}. 

\begin{algorithm}[!h]
\begin{algorithmic}[1]
\begin{footnotesize}
  \Function{Na{\"\i}ve-search}{$p$,$m$,$t$,$n$}
  \State $\textit{count} \gets 0$
  \For{$i \gets 1$ to $n-m+1$}
    \State $\textit{found} \gets \textrm{true}$
    \For{$j \gets 1 .. m$} 
      \State $\textit{found} \gets \textit{found}$ and ($t[i+j-1]=p[j]$)
      \If{$\textit{found}=\textrm{false}$}
      \State \Goto{NaiveSearch2-out} \Comment{Start checking the next position in text $t$}
    \EndIf
    \EndFor
  \State $\textit{count} \gets \textit{count} + 1$\label{NaiveSearch2-printi}
  \EndFor\Comment{destination for \textbf{go to}} \label{NaiveSearch2-out}
\EndFunction
\end{footnotesize}
\end{algorithmic}
\caption{}
\label{NaiveSearch2}
\end{algorithm}

\begin{figure}[!h]
  \centering
  \includegraphics[width=0.9\textwidth]{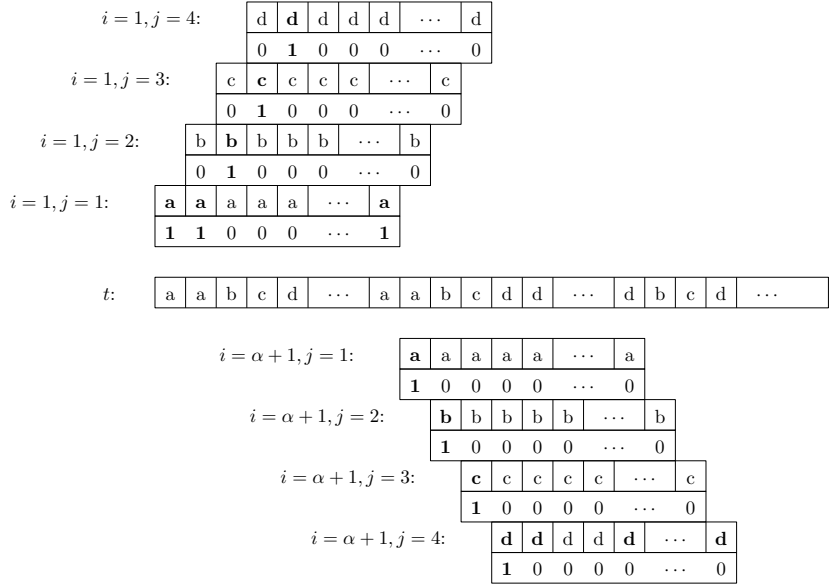}
  \caption{Example of comparisons for the text $t=\textrm{aabcd}\cdots$ and pattern $p=\textrm{abcd}$ using the \textsc{SIMD-Na{\"\i}ve-search} algorithm (alignment of pattern vector and vector \textit{found} to text $t$).}
  \label{pic@SIMDmatch}
\end{figure}
 
\begin{algorithm}[!h]
\begin{algorithmic}[1]
\begin{footnotesize}
  \Function{SIMD-Na{\"\i}ve-search}{$p$,$m$,$t$,$n$,$\alpha$}
  \State $\textit{count} \gets 0$
  \For{$i \gets 1$ to $n-m+1$ step $\alpha$}
    \State $\textit{found} \gets 11..1$
    \For{$j \gets 1 .. m$} 
      \State $\textit{found} \gets \textit{found} \textrm{ AND } \textrm{SIMDcompare}(t,i+j-1,p,j,\alpha)$
      \If{$\textit{found}=0$}
      \State \Goto{SIMDNaiveSearch2-out} \Comment{Start checking the next positions in text $t$}
    \EndIf
    \EndFor
    \State $\textit{count} \gets \textit{count} + \textrm{popcount}(\textit{found})$\label{SIMDNaiveSearch2-printi}
  \EndFor \Comment{destination for \textbf{go to}} \label{SIMDNaiveSearch2-out}
  \EndFunction
\end{footnotesize}
\end{algorithmic}
\caption{}
\label{SIMDNaiveSearch2}
\end{algorithm}

Using SIMD instructions (shown in Algorithm~\ref{SIMDNaiveSearch2}) we can compare $\alpha$ bytes in parallel, where $\alpha=16$ in case of SSE2 or $\alpha=32$ in case of AVX2 and `AND' represents the bit-parallel `and'. This allows huge speedup of a run.

For a given position $i$ in the text $t$, the idea is to compare the pattern $p$ with the $\alpha$ substrings $t[i+k..i+k+m-1]$, for $0 \leq k < \alpha$, in parallel, in ${\cal O}(m)$ time in total. To this end, we use a primitive $\textrm{SIMDcompare}(t, i, p, j, \alpha)$ which, given a position $i$ in $t$ and $j$ in $p$, compares the strings $S_1 = t[i + j - 1 .. i + j - 1 + \alpha]$ and $S_2 = p[j]^\alpha$ and returns an $\alpha$-bit integer such that the $k$-th bit is set iff $S_1[k] = S_2[k]$, in ${\cal O}(1)$ time. In other words, the output integer encodes the result of the $j$-th symbol comparison for all the $\alpha$ substrings. For example, consider the $\alpha$ leftmost substrings of length $m$ of $t$, corresponding to $i = 1$. For $j = 1$, the function compares $t[1..\alpha]$ with $p[1]^\alpha$, i.e., the first symbol of the substrings against $p[1]$. For $j = 2$, the function compares $t[2..\alpha+1]$ with $p[2]^\alpha$, i.e., the second symbol against $p[2]$.  Let \textit{found} be the bitwise and of the integers $\textrm{SIMDcompare}(t, i, p, j, \alpha)$, for $j = 1, \ldots, m$. Clearly, $t[i+k..i+k+m-1] = p$ iff the $k$-bit of \textit{found} is set. We compute \textit{found} iteratively, until we either compare the last symbol of $p$ or no substring has a partial match (i.e., the vector \textit{found} becomes zero). Then, the text is advanced by $\alpha$ positions and the process is repeated starting at position $i + \alpha$. For a given $i$, the number of occurrences of $P$ is equal to the number of bits set in \textit{found} and is computed using a popcount instruction. Reporting all matches in line~\ref{SIMDNaiveSearch2-printi} would add an ${\cal O}(s)$ time overhead, as ${\cal O}(s)$ instructions are needed to extract the positions of the bits set in \textit{found}, where $s$ is the number of occurrences found.

The 16-byte version of function SIMDcompare is implemented with SSE2 intrinsic functions as follows:

\begin{verbatim}
    SIMDcompare(x, y, 16)
        x_ptr = _mm_loadu_si128(x)
        y_ptr = _mm_loadu_si128(s(y,16))
        return _mm_movemask_epi8(_mm_cmpeq_epi8(x_ptr, y_ptr))
\end{verbatim}

\noindent Here {\tt s(y,16)} is the starting address of 16 copies of {\tt y}. The instruction \verb|_mm_loadu_si128(x)| loads 16 bytes (=128 bits) starting from {\tt x} to a SIMD register. The instruction \verb|_mm_cmpeq_epi8| compares bytewise two registers and the instruction \verb|_mm_movemask_epi8| extracts the comparison result as a 16-bit integer. For the 32-byte version, the corresponding AVX2 intrinsic functions are used. For both versions the SSE4 instruction \verb|_mm_popcnt_u32| is utilized for popcount.

\subsection{Frequency Involved}
\label{sec@Frequency_Involved}

In order to identify nonmatching positions in the text as fast as possible, individual characters of the pattern are compared to the corresponding positions in the text in the order given by their frequency in standard text. First, the least frequent symbol is compared, then the second least frequent symbol, etc. Therefore the text type should be considered and frequencies of symbols in the text type should be computed in advance from some relevant corpus of texts of the same type. Hume and Sunday \cite{HS91} use this strategy in the context of the Boyer--Moore algorithm.

\begin{algorithm}[!h]
\begin{algorithmic}[1]
\begin{footnotesize}
  \Function{Freq-SIMD-Na{\"\i}ve-search}{$p$,$m$,$t$,$n$,$\alpha$}
  \State $\textit{count} \gets 0$
  \For{$i \gets 1$ to $n-m+1$ step $\alpha$}
    \State $\textit{found} \gets 11..1$
    \For{$j \gets 1 .. m$} 
      \State $\textit{found} \gets \textit{found} \textrm{ AND } \textrm{SIMDcompare}(t, i+\pi(j)-1, p, \pi(j),\alpha)$
      \If{$\textit{found}=0$}
      \State \Goto{FreqSIMDNaiveSearch2-out} \Comment{Start checking the next positions in text $t$}
    \EndIf
    \EndFor
    \State $\textit{count} \gets \textit{count} + \textrm{popcount}(\textit{found})$\label{FreqSSIMDNaiveSearch2-printi}
  \EndFor \Comment{destination for \textbf{go to}} \label{FreqSIMDNaiveSearch2-out}
  \EndFunction
\end{footnotesize}
\end{algorithmic}
\caption{}
\label{FreqSIMDNaiveSearch2}
\end{algorithm}

Algorithm~\ref{FreqSIMDNaiveSearch2} shows the \naive{} approach enriched by frequency consideration. A function $\pi$ gives the order in which the symbols of pattern should be compared (i.e., $p[\pi(1)], p[\pi(2)],\ldots,p[\pi(m)]$) to the corresponding symbols in text. An array for the function $\pi$ is computed in ${\cal O}(m\log m)$ time using a standard sorting algorithm on frequencies of symbols in $p$.

Hume and Sunday \cite{HS91} call this strategy {\em optimal match}, although it is not necessarily optimal. For example, the pattern `qui' is tested in the order `q'-`u'-`i', but the order `q'-`i'-`u' is clearly better in practice because `q' and `u' appear often together. K\"{u}lekci \cite{Kul2007} compares optimal match with more advanced strategies based on frequencies of discontinuous $q$-grams\footnote{Basically if a symbol $p[i]$ in a position $i$ of the pattern $p$ matches to the text, compare next the position of $p$ that most unlikely matches.} with conditional probabilities. His experiments show that the frequency is beneficial in case of texts of large alphabets like texts of natural language. Computing all possible frequencies of $q$-grams is rather complicated and the possible speed-up to optimal match is likely marginal. Thus we consider only simple frequencies of individual symbols. 

\subsection{Loop Peeling}

Guard test \cite{HS91,Rai99} is a widely used technique to speed-up string matching. The idea is to test a certain pattern position before entering a checking loop. Instead of a single guard test, two or even three tests have been used \cite{Raita92,TVS06}. Guard test is a representative of a general optimization technique called loop peeling, where a number of iterations is moved in front of the loop. As a result, the loop becomes faster because of fewer loop tests. Moreover, loop peeling makes possible to precompute certain values used in the moved iterations. For example, $p[\pi(1)]$ is explicitly known. In some cases, loop peeling may even double the speed of a string matching algorithm applying SIMD computation as observed by Chhabra et al.~\cite{CFKT16}.

In the following, we call the number of the moved iterations the peeling factor $r$. We assume that the first loop test is done after $r$ iterations. Thus our approach differs from multiple guard test, where checking is stopped after the first mismatch. All $r$ iterations are performed in our approach. 

\begin{algorithm}[!h]
\begin{algorithmic}[1]
\begin{footnotesize}
  \Function{LP-Freq-SIMD-Na{\"\i}ve-search}{$p$,$m$,$t$,$n$,$\alpha$}
  \State $\textit{count} \gets 0$
  \For{$i \gets 1$ to $n-m+1$ step $\alpha$}
    \State $\textit{found} \gets \textrm{SIMDcompare}(t, i+\pi(1)-1,p, \pi(1),\alpha)$ \label{LPFreqSIMDNaiveSearch2-firstcomparison}
    \State $\textit{found} \gets \textit{found} \textrm{ AND } \textrm{SIMDcompare}(t, i+\pi(2)-1,p, p, \pi(2),\alpha)$
    \If{$\textit{found}=0$}
      \State \Goto{LPFreqSIMDNaiveSearch2-out} \Comment{Start checking the next positions in text $t$}
    \EndIf
    \For{$j \gets 3 .. m$} 
      \State $\textit{found} \gets \textit{found} \textrm{ AND } \textrm{SIMDcompare}(t, i+\pi(j)-1, p, \pi(j),\alpha)$
      \If{$\textit{found}=0$}
        \State \Goto{LPFreqSIMDNaiveSearch2-out} \Comment{Start checking the next positions in text $t$}
      \EndIf
    \EndFor
    \State $\textit{count} \gets \textit{count} + \textrm{popcount}(\textit{found})$\label{LPFreqSSIMDNaiveSearch2-printi}
  \EndFor \Comment{destination for \textbf{go to}} \label{LPFreqSIMDNaiveSearch2-out}
  \EndFunction
\end{footnotesize}
\end{algorithmic}
\caption{}
\label{LPFreqSIMDNaiveSearch2}
\end{algorithm}

Loop peeling for $r=2$ is shown in Algorithm~\ref{LPFreqSIMDNaiveSearch2}. The first two comparisons of characters are performed regardless the result of the first comparison (in line~\ref{LPFreqSIMDNaiveSearch2-firstcomparison}).

If we consider string matching in English texts, it is less probable that all the $\alpha$ comparisons fail at the same time than the other way round in the case of a pattern picked randomly from the text. Therefore it is advantageous to use the value $r=2$ for English.

In theory, $r=3$ would be good for DNA. Namely, every iteration nullifies roughly 3/4 of the remaining set bits of the bitvector \textit{found}. However, we achieved the best running time in practice with $r=5$.

\subsection{Alternative Checking Orders}
\label{sec@Alternative_Checking_Orders}

If the computation of character frequencies is considered inappropriate, there are other possibilities to speed-up checking. In natural languages adjacent characters have positive correlation. To break correlations one can use a fixed order which avoids adjacent characters. We applied the following heuristic order $\pi_h$: $p[1],p[m], p[4], p[7],\ldots, p[3], p[6], \ldots, p[2], p[5],\ldots$.

In letter-based languages, the space character is the most frequent character. We can transform $\pi_h$ to a slightly  better scheme $\pi_{hs}$ by moving first all the spaces to the end and then processing the remaining positions as for $\pi_h$.

\section{Experiments}

\begin{figure}[htb]
  \centering
  \includegraphics[width=0.9\textwidth]{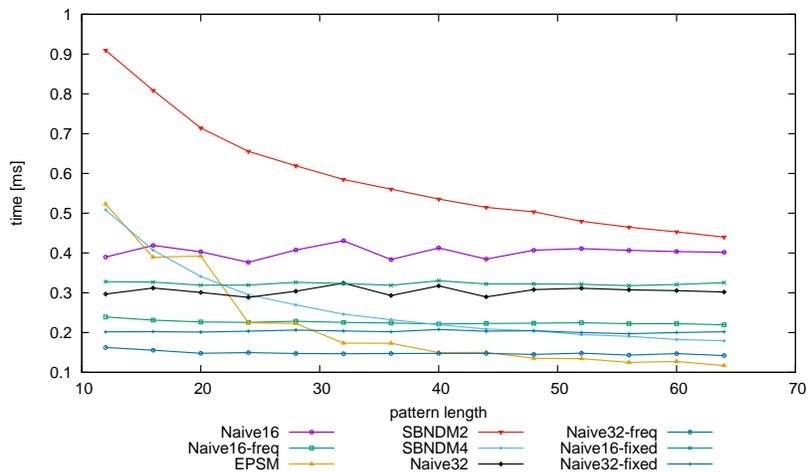}
  \caption{Search time for \texttt{bible.txt} ($|\Sigma|=63$)}
  \label{FigBible}
\end{figure}

\begin{figure}[htb]
  \centering
  \includegraphics[width=0.9\textwidth]{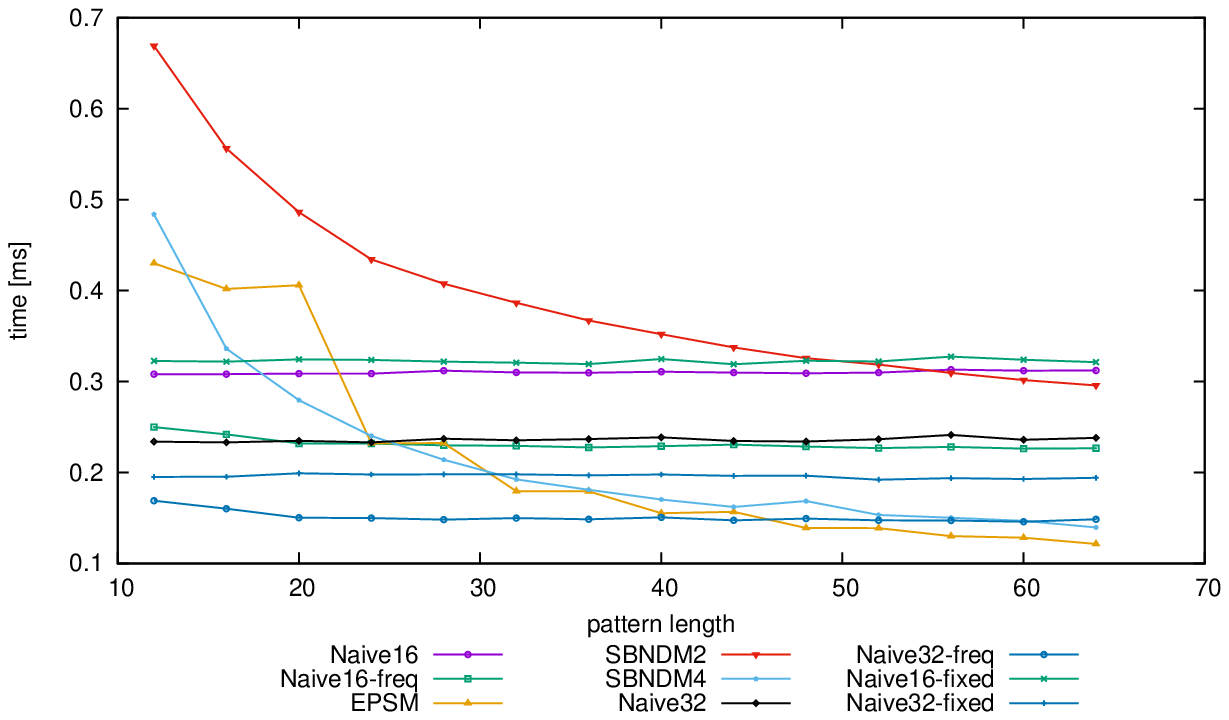}
  \caption{Search time for \texttt{Dostoevsky-TheDouble.txt} ($|\Sigma|=117$)}
  \label{FigDostoyevsky}
\end{figure}

\begin{figure}[htb]
  \centering
  \includegraphics[width=0.9\textwidth]{{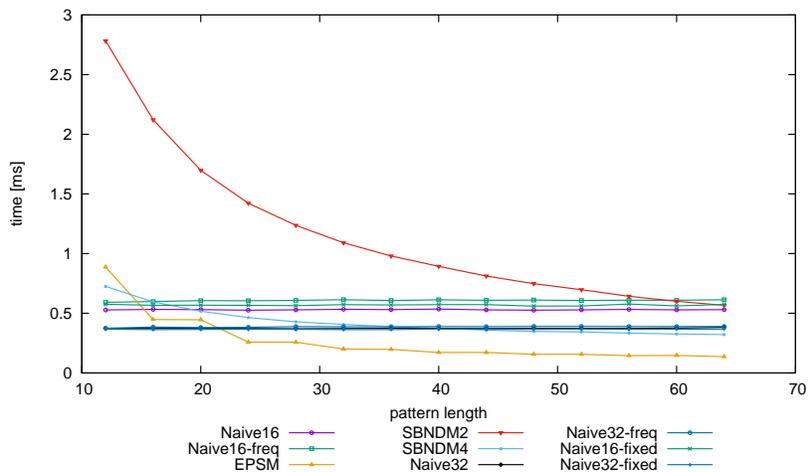}}
  \caption{Search time for \texttt{E.coli.txt} ($|\Sigma|=4$)}
  \label{FigEcoli}
\end{figure}

\begin{figure}[htb]
  \centering
  \includegraphics[width=0.9\textwidth]{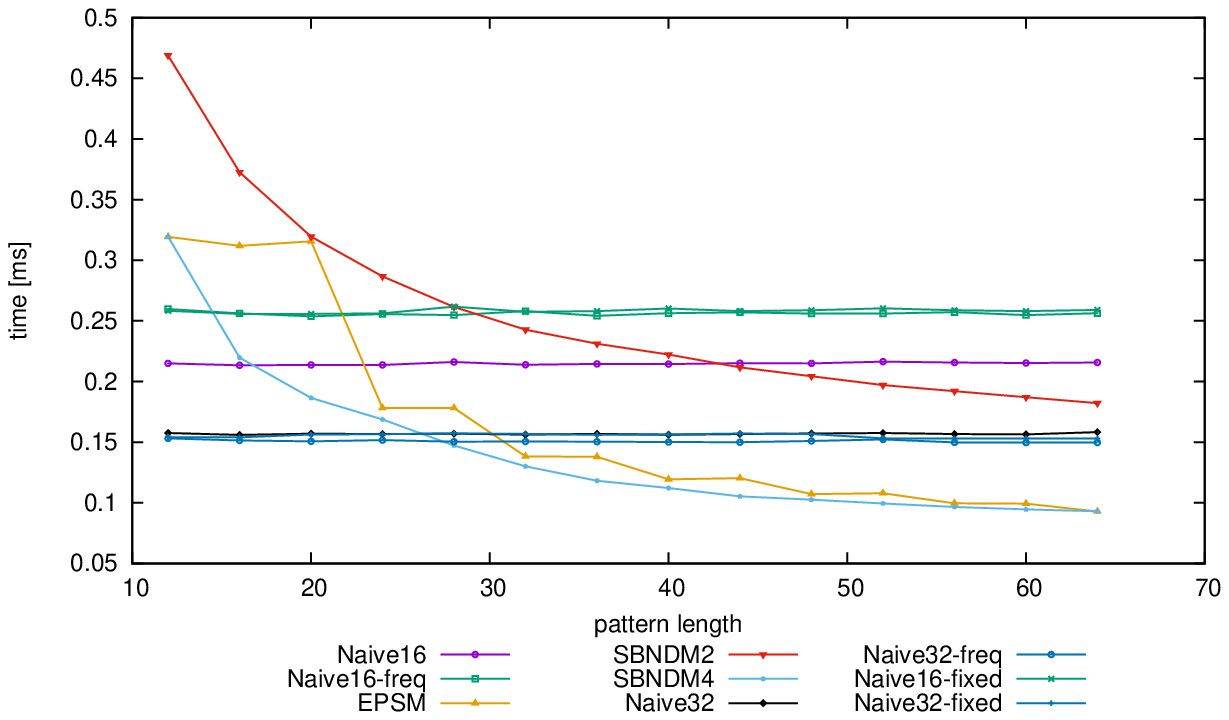}
  \caption{Search time for \texttt{plot-protein-hs.txt} ($|\Sigma|=19$)}
  \label{FigProtein}
\end{figure}

We have selected four files of different types and alphabet sizes to run experiments on: \texttt{bible.txt} (Fig.~\ref{FigBible}, Table~\ref{tab@bible}) and \texttt{E.coli.txt} (Fig.~\ref{FigEcoli}, Table~\ref{tab@Ecoli}) taken from Canterbury Corpus \cite{AB97a}, \texttt{Dostoevsky-TheDouble.txt} (Fig.~\ref{FigDostoyevsky}, Table~\ref{tab@Dostoyevsky}), novel The Double by Dostoevsky in Czech language taken from Project Gutenberg\footnote{\url{https://www.gutenberg.org/}}, and \texttt{protein-hs.txt} (Fig.~\ref{FigProtein}, Table~\ref{tab@Protein}) taken from Protein Corpus \cite{NMW99}. File \texttt{Dostoevsky-TheDouble.txt} is a concatenation of five copies of the original file to get file length similar to the other files.

\begin{table}[htb]
\scalebox{0.89}{
\begin{tabular}{r||r|r|r|r|r|r|r|r|r}
$m$     &           N16 &      N16-freq &          EPSM &        SBNDM2 &        SBNDM4 &           N32 &      N32-freq &     N16-fixed &     N32-fixed \\
  \hline
4	&	0.411	&	0.415	&	0.450	&	1.845	&	3.263	&	0.297	&	0.291	&	0.370	&\bf	0.253	\\
8	&	0.415	&	0.271	&	0.519	&	1.168	&	0.900	&	0.311	&\bf	0.190	&	0.329	&	0.214	\\
12	&	0.389	&	0.239	&	0.523	&	0.909	&	0.508	&	0.296	&\bf	0.162	&	0.327	&	0.201	\\
16	&	0.419	&	0.231	&	0.389	&	0.808	&	0.407	&	0.312	&\bf	0.155	&	0.326	&	0.202	\\
20	&	0.403	&	0.226	&	0.392	&	0.714	&	0.340	&	0.300	&\bf	0.147	&	0.319	&	0.201	\\
24	&	0.376	&	0.225	&	0.224	&	0.655	&	0.294	&	0.288	&\bf	0.149	&	0.319	&	0.203	\\
28	&	0.407	&	0.228	&	0.222	&	0.619	&	0.269	&	0.304	&\bf	0.147	&	0.326	&	0.206	\\
32	&	0.430	&	0.225	&	0.173	&	0.584	&	0.246	&	0.324	&\bf	0.146	&	0.323	&	0.203	\\
36	&	0.383	&	0.224	&	0.173	&	0.560	&	0.232	&	0.293	&\bf	0.147	&	0.318	&	0.201	\\
40	&	0.412	&	0.221	&	0.149	&	0.535	&	0.219	&	0.317	&\bf	0.147	&	0.330	&	0.207	\\
44	&	0.384	&	0.222	&	0.150	&	0.514	&	0.208	&	0.289	&\bf	0.147	&	0.322	&	0.203	\\
48	&	0.407	&	0.223	&\bf	0.135	&	0.503	&	0.204	&	0.307	&	0.145	&	0.322	&	0.204	\\
52	&	0.411	&	0.224	&\bf	0.134	&	0.479	&	0.194	&	0.311	&	0.148	&	0.321	&	0.199	\\
56	&	0.406	&	0.222	&\bf	0.124	&	0.464	&	0.190	&	0.307	&	0.143	&	0.318	&	0.197	\\
60	&	0.403	&	0.222	&\bf	0.127	&	0.453	&	0.182	&	0.305	&	0.146	&	0.320	&	0.200	\\
64	&	0.401	&	0.219	&\bf	0.116	&	0.439	&	0.179	&	0.301	&	0.142	&	0.325	&	0.201	\\
\end{tabular}
}
  \caption{Search times [ms] for \texttt{bible.txt} ($|\Sigma|=63$)}
  \label{tab@bible}
\end{table}

\begin{table}[htb]
\scalebox{0.89}{
\begin{tabular}{r||r|r|r|r|r|r|r|r|r}
$m$     &           N16 &      N16-freq &          EPSM &        SBNDM2 &        SBNDM4 &           N32 &      N32-freq &     N16-fixed &     N32-fixed \\
  \hline
4	&	0.311	&	0.306	&	0.473	&	1.780	&	3.393	&	0.232	&	0.226	&	0.330	&\bf	0.212	\\
8	&	0.307	&	0.264	&	0.439	&	0.923	&	0.775	&	0.233	&\bf	0.182	&	0.322	&	0.195	\\
12	&	0.308	&	0.249	&	0.430	&	0.669	&	0.483	&	0.233	&\bf	0.168	&	0.322	&	0.195	\\
16	&	0.308	&	0.241	&	0.401	&	0.556	&	0.336	&	0.233	&\bf	0.160	&	0.321	&	0.195	\\
20	&	0.308	&	0.232	&	0.405	&	0.486	&	0.279	&	0.234	&\bf	0.150	&	0.324	&	0.199	\\
24	&	0.308	&	0.231	&	0.232	&	0.434	&	0.240	&	0.233	&\bf	0.149	&	0.323	&	0.197	\\
28	&	0.311	&	0.229	&	0.232	&	0.407	&	0.214	&	0.237	&\bf	0.148	&	0.321	&	0.197	\\
32	&	0.309	&	0.229	&	0.179	&	0.386	&	0.192	&	0.235	&\bf	0.149	&	0.320	&	0.197	\\
36	&	0.309	&	0.227	&	0.179	&	0.367	&	0.180	&	0.236	&\bf	0.148	&	0.319	&	0.196	\\
40	&	0.310	&	0.228	&	0.155	&	0.351	&	0.170	&	0.238	&\bf	0.150	&	0.324	&	0.197	\\
44	&	0.309	&	0.230	&	0.156	&	0.337	&	0.162	&	0.234	&\bf	0.147	&	0.318	&	0.196	\\
48	&	0.309	&	0.228	&\bf	0.139	&	0.325	&	0.168	&	0.234	&	0.149	&	0.322	&	0.196	\\
52	&	0.309	&	0.226	&\bf	0.138	&	0.318	&	0.153	&	0.236	&	0.147	&	0.322	&	0.192	\\
56	&	0.313	&	0.228	&\bf	0.130	&	0.309	&	0.150	&	0.241	&	0.147	&	0.327	&	0.193	\\
60	&	0.311	&	0.226	&\bf	0.128	&	0.301	&	0.146	&	0.235	&	0.145	&	0.323	&	0.192	\\
64	&	0.312	&	0.226	&\bf	0.121	&	0.295	&	0.139	&	0.238	&	0.148	&	0.321	&	0.194	\\
\end{tabular}
}
  \caption{Search times [ms] for \texttt{Dostoevsky-TheDouble.txt} ($|\Sigma|=117$)}
  \label{tab@Dostoyevsky}
\end{table}

\begin{table}[htb]
\scalebox{0.89}{
\begin{tabular}{r||r|r|r|r|r|r|r|r|r}
$m$     &           N16 &      N16-freq &          EPSM &        SBNDM2 &        SBNDM4 &           N32 &      N32-freq &     N16-fixed &     N32-fixed \\
  \hline
4	&	0.462	&	0.651	&	0.520	&	4.749	&	3.840	&\bf	0.307	&	0.467	&	0.655	&	0.465	\\
8	&	0.526	&	0.589	&	0.890	&	3.525	&	1.017	&	0.373	&	0.375	&	0.575	&\bf	0.371	\\
12	&	0.527	&	0.592	&	0.887	&	2.781	&	0.725	&	0.373	&	0.375	&	0.576	&\bf	0.366	\\
16	&	0.533	&	0.598	&	0.447	&	2.121	&	0.595	&	0.373	&	0.384	&	0.566	&\bf	0.363	\\
20	&	0.531	&	0.606	&	0.445	&	1.698	&	0.517	&	0.373	&	0.381	&	0.567	&\bf	0.364	\\
24	&	0.525	&	0.605	&\bf	0.258	&	1.422	&	0.463	&	0.371	&	0.382	&	0.566	&	0.366	\\
28	&	0.529	&	0.608	&\bf	0.257	&	1.237	&	0.428	&	0.370	&	0.388	&	0.563	&	0.365	\\
32	&	0.533	&	0.613	&\bf	0.200	&	1.090	&	0.406	&	0.372	&	0.387	&	0.573	&	0.363	\\
36	&	0.531	&	0.606	&\bf	0.198	&	0.980	&	0.387	&	0.373	&	0.387	&	0.569	&	0.363	\\
40	&	0.535	&	0.612	&\bf	0.172	&	0.893	&	0.374	&	0.373	&	0.389	&	0.572	&	0.368	\\
44	&	0.528	&	0.608	&\bf	0.171	&	0.812	&	0.359	&	0.372	&	0.388	&	0.572	&	0.366	\\
48	&	0.526	&	0.611	&\bf	0.156	&	0.748	&	0.348	&	0.374	&	0.390	&	0.560	&	0.365	\\
52	&	0.529	&	0.607	&\bf	0.156	&	0.697	&	0.343	&	0.373	&	0.390	&	0.561	&	0.367	\\
56	&	0.532	&	0.609	&\bf	0.145	&	0.642	&	0.334	&	0.373	&	0.389	&	0.577	&	0.365	\\
60	&	0.528	&	0.609	&\bf	0.146	&	0.600	&	0.326	&	0.375	&	0.389	&	0.562	&	0.366	\\
64	&	0.530	&	0.613	&\bf	0.136	&	0.567	&	0.321	&	0.382	&	0.389	&	0.573	&	0.367	\\
\end{tabular}
}
  \caption{Search times [ms] for \texttt{E.coli.txt} ($|\Sigma|=4$)}
  \label{tab@Ecoli}
\end{table}

\begin{table}[htb]
\scalebox{0.89}{
\begin{tabular}{r||r|r|r|r|r|r|r|r|r}
$m$     &           N16 &      N16-freq &          EPSM &        SBNDM2 &        SBNDM4 &           N32 &      N32-freq &     N16-fixed &     N32-fixed \\
  \hline
4	&	0.214	&	0.254	&	0.369	&	1.219	&	2.607	&	0.156	&	0.156	&	0.254	&\bf	0.154	\\
8	&	0.214	&	0.256	&	0.319	&	0.657	&	0.531	&	0.156	&\bf	0.152	&	0.258	&	0.156	\\
12	&	0.215	&	0.259	&	0.319	&	0.469	&	0.319	&	0.157	&\bf	0.153	&	0.258	&	0.154	\\
16	&	0.213	&	0.256	&	0.311	&	0.372	&	0.219	&	0.156	&\bf	0.151	&	0.255	&	0.153	\\
20	&	0.213	&	0.253	&	0.315	&	0.319	&	0.186	&	0.157	&\bf	0.150	&	0.255	&	0.156	\\
24	&	0.213	&	0.255	&	0.178	&	0.286	&	0.168	&	0.156	&\bf	0.151	&	0.256	&	0.156	\\
28	&	0.216	&	0.254	&	0.178	&	0.261	&\bf	0.147	&	0.157	&	0.150	&	0.261	&	0.157	\\
32	&	0.213	&	0.258	&	0.138	&	0.242	&\bf	0.130	&	0.156	&	0.150	&	0.257	&	0.156	\\
36	&	0.214	&	0.254	&	0.137	&	0.231	&\bf	0.118	&	0.156	&	0.150	&	0.258	&	0.155	\\
40	&	0.214	&	0.256	&	0.119	&	0.222	&\bf	0.112	&	0.156	&	0.150	&	0.260	&	0.156	\\
44	&	0.215	&	0.257	&	0.120	&	0.211	&\bf	0.105	&	0.156	&	0.149	&	0.258	&	0.157	\\
48	&	0.214	&	0.256	&	0.107	&	0.204	&\bf	0.102	&	0.157	&	0.151	&	0.258	&	0.156	\\
52	&	0.216	&	0.256	&	0.107	&	0.197	&\bf	0.099	&	0.157	&	0.152	&	0.260	&	0.153	\\
56	&	0.215	&	0.257	&	0.099	&	0.192	&\bf	0.096	&	0.156	&	0.149	&	0.258	&	0.152	\\
60	&	0.215	&	0.254	&	0.099	&	0.187	&\bf	0.094	&	0.156	&	0.149	&	0.258	&	0.153	\\
64	&	0.215	&	0.256	&\bf	0.092	&	0.182	&	0.092	&	0.158	&	0.149	&	0.259	&	0.152	\\
\end{tabular}
}
  \caption{Search times [ms] for \texttt{plot-protein-hs.txt} ($|\Sigma|=19$)}
  \label{tab@Protein}
\end{table}

We have compared methods Naive16 and Naive32 having 16 and 32 bytes processed by one SIMD instruction respectively. Naive16-freq and Naive32-freq are their variants where comparison order given by nondecreasing probability of pattern symbols (Section~\ref{sec@Frequency_Involved}). Naive16-fixed and Naive32-fixed are the variants where comparison order is fixed (Section~\ref{sec@Alternative_Checking_Orders}). Our methods were compared with the fastest exact string matching algorithms \cite{FK2013} up to now SBNDM2, SBNDM4 \cite{DHPT2010} and EPSM\footnote{As of July 2016, the EPSM algorithm in the SMART library has problems for pattern lengths greater than 16. We used a corrected version in our tests.} \cite{FK2013} taken from SMART Library\footnote{\url{http://www.dmi.unict.it/~faro/smart/}}.

The experiments were run on GNU/Linux 3.18.12, with x86\_64 Intel\textregistered{} Core\texttrademark{} i7-4770 CPU 3.40GHz with 16GB RAM. The computer was without any other workload and user time was measured using POSIX function \texttt{getrusage()}. The average of 100 running times is reported. The accuracy of the results is about $\pm 2\%$.

The experiments show for both SSE2 and AVX2 instructions that for natural text (\texttt{bible.txt}) with the scheme $\pi_h$ of fixed frequency of comparisons improves the speed of \textsc{SIMD-Na{\"\i}ve-search} but it is further improved by considering frequencies of symbols in the text. In case of natural text with larger alphabet (\texttt{Dostoevsky-TheDouble.txt}) the scheme $\pi_h$ improves the speed only for AVX2 instructions. The comparison based on real frequency of symbols is the bext for both SSE2 and AVX2 instructions. In case of small alphabets (\texttt{E.coli.txt}, \texttt{protein-hs.txt}) the order of comparison of symbols does not play any role (except for \texttt{protein-hs.txt} and SSE2 instructions).

For files with large alphabet (\texttt{bible.txt}, \texttt{Dostoevsky-TheDouble.txt}) the peeling factor $r=3$ gave the best results for all our algorithms except for Naive16-freq and Naive32-freq where $r=2$ was the best. The smaller the alphabet is, the less selective the bigrams or trigrams are. For file \texttt{protein-hs.txt}, $r=3$ was still good and but for DNA sequences of four symbols, $r=5$ turned to be the best

We also tested \textsc{Na{\"\i}ve-search}. In every run it was naturally considerably slower than \textsc{SIMD-Na{\"\i}ve-search}. Frequency order and loop peeling can also be applied to \textsc{Na{\"\i}ve-search}. However, the speed-up was smaller than in case of \textsc{SIMD-Na{\"\i}ve-search} in our experiments.

\section{Concluding remarks} In spite of how many algorithms were developed for exact string matching, their running times are in general outperformed by the AVX2 technology. The implementation of the \naive{} search algorithm (\textsc{Freq-SIMD-Na{\"\i}ve-search}) which uses AVX2 instructions, applies loop peeling, and compares symbols in the order of increasing frequency is the best choice in general. However, previous algorithms EPSM and SBNDM4 have an advantage for small alphabets and long patterns. Short patterns of 20 characters or less are objects of most searches in practice and our algorithm is especially good for such patterns. For texts with expected equiprobable symbols (like in DNA or protein strings), our algorithm naturally works well without the frequency order of symbol comparisons. Our algorithm is considerably simpler than its SIMD-based competitor EPSM which is a combination of six algorithms.

\section*{Acknowledgemet}
This work was done while Jan Holub was visiting the Aalto University under the AScI Visitor Programme (Dean's decision 12/2016).

\bibliographystyle{plain}

\end{document}